\documentclass{article}
\usepackage{graphicx}  
\usepackage{amsmath}   
\usepackage[compress]{cite}
\usepackage{amssymb}   
\usepackage{bm} 
\usepackage{dcolumn}
\usepackage{color}
\usepackage{mathrsfs}
\usepackage{amsfonts}
\usepackage{varioref}
\usepackage{textcomp}
\usepackage{subcaption}
\RequirePackage[colorlinks,citecolor=blue,urlcolor=magenta,linkcolor=blue]{hyperref}
\allowdisplaybreaks
\labelformat{section}{Section #1} 
\labelformat{subsection}{Section #1} 
\labelformat{subsubsection}{Section #1}
\labelformat{subsubsubsection}{Section #1}
\labelformat{equation}{Eq.~(#1)} 
\labelformat{figure}{Fig.~#1} 
\labelformat{subfigure}{Fig.~\thefigure#1} 
\labelformat{table}{Tab.~#1} 
\labelformat{appendix}{Appendix #1}
\title{\bf Universality of the Buchdahl sphere}
\author{Sumanta Chakraborty\footnote{tpsc@iacs.res.in}$~^{1}$ and Naresh Dadhich\footnote{nkd@iucaa.in}$~^{2}$
\\
$^{1}${\small{School of Physical Sciences}}\\
{\small{Indian Association for the Cultivation of Science, Kolkata-700032, India}}\\
$^{2}${\small{IUCAA, Post Bag 4, Ganeshkhind, Pune 411007}}}
\begin{document}
  
\maketitle
\begin{abstract}
Buchdahl sphere, the limiting stable isotropic stellar structure without exotic matter, plays a very important role in our understanding of how compact an astrophysical object can be. Here, we show certain universal properties associated with the Buchdahl sphere, in the sense that these properties will not change with the inclusion of electric charge in the stellar structure, or, will hold good in the pure Lovelock theories of gravity as well. Using these universalities, we have proposed a Buchdahl limit for a slowly-rotating stellar configuration, for the first time. Finally, the universality of the Buchdahl sphere in terms of the gravitational and non-gravitational field energies, as well as for the photon sphere have also been discussed. 
\end{abstract}
\section{Introduction}

The study of compact objects and their stability is important from various perspectives, ranging from explaining the observed stellar structures in our universe along with the physical processes within it, to collapse of these stellar objects to neutron stars and black holes. In this connection, it is worthwhile to ask --- what is the limiting stellar configuration possible with normal matter --- before it collapses to a black hole and disappears beyond the event horizon. Such a limit was first derived by Buchdahl, with generic stability arguments about the isotropic stellar material --- in the context of four dimensional general relativity \cite{Buchdahl:1959zz}. Subsequently, the Buchdahl limit has been generalized to other theories of gravity, including those which involves higher spacetime dimensions and with various possible matter contents \cite{Mak:2001gg,Dadhich:2010qh,Giuliani:2007zza,Boehmer:2007gq,Dadhich:2016fku,Dadhich:2019jyf,Dadhich:2016wtb,Goswami:2015dma,Karageorgis:2007cy,Andreasson:2007ck,Chakraborty:2020ifg,Rosa:2020hex,Feng:2018jrh,Feng:2021acd}. We will refer to such a limiting stellar configuration as the \emph{Buchdahl sphere} and the corresponding limit on the compactness ratio will be refereed to as the \emph{Buchdahl limit}. Note that these definitions are generic and can be introduced irrespective of the theory of gravity under consideration. For four dimensional general relativity, with isotropic matter, the Buchdahl limit correspond to the following compactness ratio, $(M/R)=(4/9)$ and the Buchdahl sphere correspond to an isotropic matter distribution, whose radius is (9/4) times its mass. Since the radius of the Buchdahl sphere is within the region between the black hole horizon and the photon sphere, its importance for gravitational wave and black hole shadow measurements cannot be underestimated \cite{Cardoso:2019rvt,Maggio:2021ans,Banerjee:2019nnj,Mishra:2019trb,Mark:2017dnq,Pani:2018flj,Maselli:2017cmm,Pani:2010em,Dey:2020lhq,Mishra:2021waw,Chakraborty:2021gdf,Chakraborty:2022zlq}. Therefore, the existence of such limiting stellar structures and study of their physical properties are always encouraging in its own right.

Here we demonstrate some new exciting properties of the Buchdahl sphere. First of all, we show that the escape velocity from the Buchdahl sphere is universal, i.e., the escape velocity remains the same irrespective of whether the stellar configuration is with or, without any electric charge. This then prompts one to assess the Buchdahl limit of a slowly-rotating stellar configuration, within general relativity, for the first time. Since, we do not know of any interior solution of the Einstein's field equations, for an arbitrarily rotating stellar configuration, we will content ourselves with the slow rotation limit. To our surprise, this universality also translates into higher curvature theories of gravity, e.g., in the context of pure Lovelock theories. In addition to the escape velocity, the universal nature of the Buchdahl sphere also reflects in the expression for the gravitational potential, as well as in the partition between the Brown-York quasi-local gravitational energy and a suitably defined non-gravitational energy (for a definition, see \cite{Chakraborty:2015kva}). These universal properties of the Buchdahl sphere very much remind us of the properties of the black hole spacetimes in general relativity and beyond, see \cite{Dadhich:2022yuk}. Thus it will be worthwhile to pursue the implications of these properties of the Buchdahl sphere further, \'{a} la those of black hole spacetimes. Also note that the analysis presented here is for isotropic fluid sphere, thus following \cite{Alho:2022bki}, it will be interesting to observe whether such universal properties will hold true for anisotropic Buchdahl sphere as well. We would like to emphasize that, our analysis will be based on stellar interiors with non-uniform density distribution, so that the velocity of sound remains less than that of light, as well as the pressure remains positive and finite at the centre of the stellar structure, even when the Buchdahl limit is achieved.

The paper is organized as follows, in \ref{univ_esc_vel_einstein} we show that the escape velocity, as well as the gravitational potential, associated with the Buchdahl sphere is indeed universal in four dimensional general relativity. Then in \ref{univ_esc_vel_lovelock}, we show that the universality of the Buchdahl sphere transcends general relativity and holds in pure Lovelock theories as well, with or without electric charge. Subsequently, we apply this universality in the context of slowly rotating stellar structure in general relativity, in \ref{buchdahl_rotating}. Finally, such universal properties of the Buchdahl sphere have been extended in the context of quasi-local gravitational energy in \ref{univ_grav_energy_buchdahl} and then we conclude with discussion on our results and possible future directions. 

\emph{Notations and Conventions}: Throughout this article we have followed the mostly positive signature convention, such that the flat spacetime metric in the Cartesian coordinate system takes the form, $\eta_{\mu \nu}=\textrm{diag.}(-1,+1,+1,+1)$. Furthermore, we have taken the fundamental constants, such that $c=1=G$. 

\section{Universality of escape velocity and potential in four-dimensional Einstein gravity}\label{univ_esc_vel_einstein}

In this section we will show that the escape velocity from a Buchdahl sphere in four-dimensional general relativity is universal irrespective of the presence of electrically charged or neutral matter inside the stellar configuration. For this purpose, we will first present the escape velocity from the surface of a static and spherically symmetric star. This requires one to define notion of the velocity of a particle in a coordinate invariant manner. However, the velocity must be observer dependent, for different observer will observe different velocities for the same particle. Following \cite{Chakraborty:2012mz}, one can write down the velocity of a particle with four-velocity $u^{\mu}$ with respect to an observer with four-velocity $u_{\rm obs}^{\mu}$, which takes the form, 
\begin{align}\label{velocity_def}
v^{2}=1-\frac{1}{(u^{\rm obs}_{\mu}u^{\mu})^{2}}~.
\end{align}
For a static and spherically symmetric spacetime, the metric describing the spacetime geometry outside the stellar structure can be expressed as, $g_{\mu \nu}=\textrm{diag.}(-f(r),\{1/f(r)\},r^{2},r^{2}\sin^{2}\theta)$. Since we are interested in the escape velocity, we take $u^{\mu}$ to be the four velocity of an outgoing geodesic and $u_{\rm obs}^{\mu}$ is taken to be a static observer at radius $r$, determining the velocity of the outgoing geodesic at radius $r$.  Thus for the static observer at radius $r$, the four-velocity is given by, $u_{\mu}^{\rm obs}=(\sqrt{f(r)},0,0,0)$, while for the timelike outgoing geodesic, $u^{0}=\{E/f(r)\}$, where $E$ is the energy of the particle. Thus, we obtain $u^{\mu}u_{\mu}^{\rm obs}=(E/\sqrt{f(r)})$. Therefore, according to \ref{velocity_def}, the velocity of the outgoing geodesic with respect to a static observer is given by, 
\begin{align}\label{velocity_exp}
v^{2}=1-\frac{f(r)}{E^{2}}~.
\end{align}
The above expression can also be derived along the following lines --- first of all from the line element of static and spherically symmetric spacetime we obtain, $d\tau^{2}=f(r)dt^{2}-\{dr^{2}/f(r)\}-r^{2}d\Omega^{2}$. For motion on the equatorial plane, we can rewrite this expression as, $1=f(r)\dot{t}^{2}(1-v^{2})$, such that $v^{2}=1-(1/f(r)\dot{t}^{2})$. For outgoing geodesic, it follows that, $\dot{t}=\{E/f(r)\}$ and hence \ref{velocity_exp} immediately follows. In order to derive the escape velocity, it is necessary that the particle should just be able to reach infinity, i.e., the particle should be at rest at infinity. Thus setting $v=0$, as $f(r)\rightarrow 1$, in \ref{velocity_exp}, we obtain $E=1$. Thus the escape velocity of a particle from a radius $r$, in a static and spherically symmetric spacetime is given by, 
\begin{align}\label{escape_final}
v_{\rm esc}^{2}=1-f(r)~.
\end{align}
This is the expression for the escape velocity $v_{\rm esc}$ we will use in order to derive its universal bound for Buchdahl sphere. In what follows we will use the Schwarzschild as well as the Reissner-Nordstr\"{o}m spacetime in order to determine the escape velocity from the surface of a spherical star with or without electric charge.

\subsection{Schwarzschild spacetime}

Let us consider the spacetime outside a massive static and spherically symmetric star, which is given by the Schwarzschild solution. For which the unknown function $f(r)$ appearing in the line element takes the form, $f(r)=1-(2M/r)$ and hence from \ref{escape_final} the escape velocity of a massive particle from the surface of the star with radius $R$ becomes,
\begin{align}\label{escape_vac}
v_{\rm esc}^{2}=\frac{2M}{R}~.
\end{align}
Note that the above escape velocity coincides with the corresponding Newtonian expression. If we now consider the star to be on the limit of stability, i.e., the radius of the star satisfying the Buchdahl limit, with $R=(9/8)(2M)$, it follows that, 
\begin{align}
v^{2}_{\rm limit}=\frac{8}{9}~.
\end{align}
This provides the limiting escape velocity from the surface of a massive static and spherically symmetric star. Another interesting feature is the gravitational potential at the surface of the Buchdahl sphere. In general the gravitational potential $\Phi$ at the surface of a star with radius $R$ is given by, $\Phi=-(M/R)$, such that on the Buchdahl sphere the gravitational potential becomes, $\Phi_{\rm buchdahl}=-(4/9)$. We will now demonstrate that the same limiting velocity and gravitational potential applies even in the presence of the electric charge.  

\subsection{Reissner-Nordstr\"{o}m spacetime}

Let us now consider the case of a charged, static and spherically symmetric shell with mass $M$ and electric charge $Q$. The exterior geometry is given by the Reissner-Nordstr\"{o}m spacetime, for which the unknown function $f(r)$ takes the form, $f(r)=1-(2M/r)+(Q^{2}/r^{2})$. In which case, the escape velocity from the surface of this spherical shell (with radius $R$) follows from \ref{escape_final} and is given by 
\begin{align}\label{escape_RN}
v^{2}_{\rm esc}=\frac{2M}{R}-\alpha^{2}\frac{M^{2}}{R^{2}}~;\qquad \alpha\equiv \frac{Q}{M}~.
\end{align}
Note that for $\alpha=0$, the above escape velocity reduces to the corresponding expression for Schwarzschild spacetime. If we now consider the radius of this charged shell to coincide with the corresponding Buchdahl limit (such a shell, whose compactness coincides with the respective Buchdahl limit, is referred to as the \emph{Buchdahl shell}) , which takes the form \cite{Chakraborty:2020ifg,Boehmer:2007gq,Giuliani:2007zza,Andreasson:2007ck,Andreasson:2012dj},
\begin{align}\label{Buchdahl_charged}
\frac{M}{R}=\frac{8}{9}\left[1+\sqrt{1-\frac{8}{9}\alpha^{2}}\right]^{-1}~,
\end{align}
then from \ref{escape_RN} the limiting form of the escape velocity from its surface becomes,
\begin{align}
v^{2}_{\rm limit}&=\frac{16}{9}\left[1+\sqrt{1-\frac{8}{9}\alpha^{2}}\right]^{-1}-\alpha^{2}\left(\frac{8}{9}\right)^{2}\left[1+\sqrt{1-\frac{8}{9}\alpha^{2}}\right]^{-2}
\nonumber
\\
&=\frac{8}{9}\left[1+\sqrt{1-\frac{8}{9}\alpha^{2}}\right]^{-2}\left\{2\left[1+\sqrt{1-\frac{8}{9}\alpha^{2}}\right]-\frac{8}{9}\alpha^{2}\right\}=\frac{8}{9}~.
\end{align}
Therefore, we observe that whether the exterior geometry is Schwarzschild or, Reissner-Nordstr\"{o}m, the escape velocity from the limiting stellar configuration is always the same, given by $v_{\rm limit}^{2}=(8/9)$. This suggests that the appearance of a limiting radius for stellar structure is intertwined with the existence of a universal escape velocity for four dimensional general relativity. 

Note that for $\alpha^{2}=(9/8)$, the Buchdahl limit becomes, $(M/R)=(8/9)$, which is twice of the (M/R) ratio for a Buchdahl sphere in vacuum. This is identical to the extremal black hole scenario, where the horizon radius of an extremal Reissner-Nordstr\"{o}m black hole is precisely twice of the horizon radius for Schwarzschild black hole. Moreover, the above suggests that for $1<(Q/M)<(3/2\sqrt{2})$, there can be stable stellar structure which cannot collapse to a black hole, but possibly to a naked singularity. Thereby providing a formation channel for naked singular spacetimes. Finally, in the present scenario, the gravitational potential at the surface of a spherical object with radius $R$ reads, $\Phi=-(M/R)+(Q^{2}/2R^{2})$, such that in the Buchdahl limit, this gravitational potential becomes, $\Phi_{\rm buchdahl}=-2v_{\rm limit}^{2}=-(4/9)$, identical to that of the Schwarzschild black hole. This shows that both the escape velocity and the gravitational potential is universal for Buchdahl sphere with or without charge. We will now demonstrate that this result transcends general relativity and holds true for higher curvature theories as well, which will put forward the universality of the escape velocity from the limiting stellar configuration to a firm ground and then we will use the same subsequently, in obtaining the buchdahl limit for rotating star as well.  

\section{Universality of escape velocity and potential in pure Lovelock theory of gravity}\label{univ_esc_vel_lovelock}

The above analysis motivates one to examine if such a universal nature of the escape velocity and the gravitational potential exists beyond general relativity as well. For this purpose we will discuss if such universal behaviour transcends general relativity. Following which, we will study the escape velocity from Buchdahl sphere and the gravitational potential on the surface of the Buchdahl sphere, with or, without the electric charge in $d$ spacetime dimensions for pure Lovelock theory of order $N$. As in general relativity, we will first present the case of vacuum solutions before discussing the charged case.   

Let us briefly describe the structure of the pure Lovelock theory. The Lovelock theories of gravity has a special place within the domain of higher curvature theories of gravity, since despite having higher curvature terms, these theories yield second order gravitational field equation, thereby exorcising the Ostrogradsky instability. In general, the Lovelock Lagrangian in $d$ spacetime dimensions involves --- $k\leq (d/2)$ number of terms, for even spacetime dimensions $d$, and for odd number of spacetime dimensions, the number of terms in the Lovelock Lagrangian corresponds to, $k\leq (d-1)/2$. Each of these terms is a polynomial involving the Riemann tensor, such that the $N$th term in the series involves $N$ Riemann tensors appropriately contracted with the alternative tensor. Written explicitly, the full Lovelock Lagrangian in $d$ spacetime dimensions, take the following form, 
\begin{align}
L=\sum_{N=1}^{k}\frac{1}{2^{N}}\delta^{\alpha_{1}\beta_{1}\alpha_{2}\beta_{2}\cdots \alpha_{N}\beta_{N}}_{\mu_{1}\nu_{1}\mu_{2}\nu_{2}\cdots \mu_{N}\nu_{N}}R^{\mu_{1}\nu_{1}}_{\alpha_{1}\beta_{1}}R^{\mu_{2}\nu_{2}}_{\alpha_{2}\beta_{2}}\cdots R^{\mu_{N}\nu_{N}}_{\alpha_{N}\beta_{N}}~.
\end{align}
The pure Lovelock theories of gravity, on the other hand, corresponds to a single term in this series. For a pure Lovelock theory of order $N$, we have $N$ Riemann tensors combined together with an alternative tensor of appropriate rank. Such a pure Lovelock theory of order $N$ is dynamical in spacetime dimensions $d\geq 2N+2$. This is the scenario considered below.  

At this outset, let us discuss the issue of the initial value problem and causality constraints for Lovelock theories of gravity. First of all, even though it is true that gravitons in Lovelock theories can move faster than light, this cannot be used to construct closed timelike curves, as advocated in \cite{Camanho:2014apa}, since such scenarios can never arise from a regular initial data \cite{Papallo:2015rna}. The field equations for Lovelock gravity, on the other hand, does not admit even weak hyperbolicity --- necessary for the well-posedness of the initial value problem --- if the standard gauge choices, as done in the case of general relativity, are considered \cite{Papallo:2017qvl}. However, it turns out that there are modified gauge choices, for which the Lovelock field equations are strongly hyperbolic \cite{Kovacs:2020ywu}. This suggest that the initial value problem is indeed well-posed. 

\subsection{Exterior vacuum spacetime}

Let us consider the case of a static and spherically symmetric stellar configuration with vacuum exterior. Since the gravitational interaction is considered to be described by pure Lovelock gravity of order $N$ in $d$ spacetime dimensions, the unknown function $f(r)$ appearing in the diagonal metric $g_{\mu \nu}=\textrm{diag.}(-f(r),1/f(r),r^{2},r^{2}\sin^{2}\theta,\cdots)$ read, $f(r)=1-(2^{N}M/r^{d-2N-1})^{1/N}$ \cite{Dadhich:2012pd,Dadhich:2010gu,Dadhich:2012ma,Chakraborty:2016qbw}. Note that for $N=1$ and $d=4$, the function $f(r)$ becomes identical to that of the Schwarzschild spacetime. Therefore from \ref{escape_final}, the escape velocity of a particle from the surface of the star with radius $R$, within the pure Lovelock theories of gravity reads,  
\begin{align}\label{escape_lovelock}
v^{2}_{\rm esc}=\left(\frac{2^{N}M}{R^{d-2N-1}}\right)^{1/N}~.
\end{align}
It is to be noted that, for $N=1$ and $d=4$, which is the four dimensional general relativity, the above expression for escape velocity reduces to the one presented in \ref{escape_vac}, as expected. The mass and radius of the Buchdahl sphere for the $N$th order Lovelock gravity in $d$ spacetime dimensions, on the other hand, are related by the following relation \cite{Dadhich:2016fku}, 
\begin{align}
M^{1/N}=\frac{2N(d-N-1)}{(d-1)^{2}}R^{\frac{d-2N-1}{N}}~,
\end{align}
whose substitution in the expression for the escape velocity in \ref{escape_lovelock} yields the following expression for the limiting velocity from the surface of the Buchdahl sphere, 
\begin{align}\label{limit_vel_vac}
v^{2}_{\rm limit}=\frac{4N(d-N-1)}{(d-1)^{2}}~.
\end{align}
This limiting velocity is independent of the mass and hence is an universal feature, depending alone on the order of the Lovelock Lagrangian and the spacetime dimensions. Further, the gravitational potential on a sphere of radius $R$ in pure Lovelock gravity reads, $\Phi=-(M/R^{d-2N-1})^{1/N}$, such that on the Buchdahl sphere, the gravitational potential reads, $\Phi_{\rm Buchdahl}=-2N(d-N-1)/(d-1)^{2}$. Note that for $N=1$ and $d=4$, the above expression for limiting velocity becomes $(8/9)$, the universal velocity for four dimensional general relativity. Also the gravitational potential on the Buchdahl sphere reduces to $-(4/9)$, as expected. Moreover, for $d=3N+1$, we obtain, $v_{\rm limit}^{2}=(8/9)$ and $\Phi_{\rm Buchdahl}=-(4/9)$, which once again demonstrates the special role played by this spacetime dimension for pure Lovelock theory of order $N$, as the results are consistent with those of four dimensional general relativity (see \cite{Chakraborty:2016qbw,Dadhich:2019jyf,Dadhich:2016fku} for a few more similar scenario).  

\subsection{Charged solution}

To see if the above expression for limiting velocity, as in \ref{limit_vel_vac}, holds true in the presence of electric charge as well, similar to the case of general relativity, here also we consider a static and spherically symmetric Buchdahl charged shell. For the charged exterior solution, the unknown function $f(r)$ becomes, $f(r)=1-\{(2^{N}M/r^{d-2N-1})-(Q^{2}/r^{2d-2N-4})\}^{1/N}$, such that using \ref{escape_final} the escape velocity from the surface of the charged shell with radius $R$ takes the form,
\begin{align}
v^{2}_{\rm esc}=\left(\frac{2^{N}M}{R^{d-2N-1}}-\frac{Q^{2}}{R^{2d-2N-4}}\right)^{1/N}
\end{align}
The Buchdahl limit, on the other hand, provides the following condition between the mass $M$, radius $R$ and electric charge $Q$ of the spherical shell \cite{Chakraborty:2020ifg}, 
\begin{align}
\left[\left(\frac{M}{R^{d-2N-1}}\right)-\frac{1}{2^{N}}\left(\frac{Q^{2}}{R^{2d-2N-4}}\right)\right]^{1/N}=\frac{2N(d-N-1)}{(d-1)^{2}}~.
\end{align}
Substitution of the same in the expression for escape velocity yields the following limiting expression for escape velocity,
\begin{align}
v^{2}_{\rm limit}=\frac{4N(d-N-1)}{(d-1)^{2}}~,
\end{align}
which coincides identically with the corresponding one derived for the vacuum exterior. Therefore, for a given order of the Lovelock polynomial and spacetime dimensions, the escape velocity from the limiting stellar configuration is universal, i.e., independent of the presence of electric charge. Moreover, for any lovelock theory of gravity of order $N$, in the spacetime dimension $d=3N+1$, the limiting velocity reads $v^{2}_{\rm limit}=(8/9)$. Thus the existence of a universal escape velocity from a Buchdahl sphere transcends general relativity. The same limiting velocity holds true for pure Gauss-Bonnet gravity in seven dimensions and so on. Finally, the gravitational potential, which for the charged case is given by, $\Phi=-\{(M/R^{d-2N-1})-(1/2^{N})(Q^{2}/R^{2d-2N-4})\}^{1/N}$ and on the Buchdahl sphere this equals $\Phi_{\rm Buchdahl}=-2N(d-N-1)/(d-1)^{2}$ --- identical to that of the vacuum spacetime considered above. Thus the universality of the gravitational potential also transcends general relativity and holds in pure Lovelock theories of gravity as well. We will now use this universal escape velocity in order to determine the Buchdahl limit for rotating stars in general relativity. 
\section{Buchdahl bound for rotating star}\label{buchdahl_rotating}

The Buchdahl bound for a rotating star, using the gravitational field equations in its interior, is very difficult to arrive at, since the field equations with matter in the presence of axi-symmetry is enormously complicated. Fortunately, the present analysis relating the existence of a limiting universal escape velocity with the Buchdahl sphere, suggests that the Buchdahl bound for rotating objects, can be obtained through the determination of the escape velocity in such spacetimes. In this section, we will first present a general expression for the escape velocity in a rotating spacetime, which we will use subsequently to find out the Buchdahl bound in the slowly rotating limit.

For axi-symmetry and stationary spacetimes, there are five metric functions that govern the spacetime geometry. We assume that these functions are such that the motion on the equatorial plane, located at $\theta=(\pi/2)$, is possible. For such a scenario, the trajectory of a particle on the equatorial plane in the rotating spacetime satisfies the following condition,
\begin{align}\label{gen_metric}
d\tau^{2}=\left(e^{2\nu}-\omega^{2}e^{2\psi}\right)dt^{2}+2\omega e^{2\psi}dt d\phi-e^{2\mu_{2}}dr^{2}-e^{2\psi}d\phi^{2}~,
\end{align}
where, $e^{2\nu}$, $\omega$, $e^{2\psi}$ and $e^{2\mu_{2}}$ are the four unknown functions governing the geometry of the equatorial plane. In general these are all functions of the radial coordinate $r$ and angular coordinates $\theta$, however for the equatorial motion, these are functions of the radial coordinate alone. In order to obtain the escape velocity, we re-write the above expression as,
\begin{align}\label{vel_rot}
d\tau^{2}\equiv\left(e^{2\nu}-\omega^{2}e^{2\psi}\right)dt^{2}\left[1-v^{2}\right]~,
\end{align}
from which the velocity of a massive particle particle moving on the equatorial plane takes the following form,
\begin{align}
v^{2}=\frac{e^{2\mu_{2}}}{\left(e^{2\nu}-\omega^{2}e^{2\psi}\right)}\left(\frac{dr}{dt}\right)^{2}+\frac{e^{2\psi}}{\left(e^{2\nu}-\omega^{2}e^{2\psi}\right)}\left(\frac{d\phi}{dt}\right)^{2}-\frac{2\omega e^{2\psi}}{\left(e^{2\nu}-\omega^{2}e^{2\psi}\right)}\left(\frac{d\phi}{dt}\right)~.
\end{align}
However, this expression for the velocity of a massive particle on the equatorial plane is not so useful for our purpose. Rather, It is possible to derive an alternative expression for the velocity from \ref{vel_rot}, which can be expressed in terms of $(dt/d\tau)$ as,
\begin{align}\label{velocity_alternative}
v^{2}=1-\frac{1}{\left(e^{2\nu}-\omega^{2}e^{2\psi}\right)\left(\frac{dt}{d\tau}\right)^{2}}~.
\end{align}
The quantity $(dt/d\tau)$ can be expressed in terms of the conserved energy $E$ (the metric elements are independent of time $t$) and conserved angular momentum $L$ (the metric elements are independent of angular coordinate $\phi$), per unit mass, such that,
\begin{align}
\frac{dt}{d\tau}=e^{-2\nu}\left(E-\omega L\right)~.
\end{align}
Substitution of the above expression for $(dt/d\tau)$ in the velocity expression as in \ref{velocity_alternative} yields, 
\begin{align}\label{escape_rotate}
v^{2}=1-\frac{e^{4\nu}}{\left(e^{2\nu}-\omega^{2}e^{2\psi}\right)\left(E-\omega L\right)^{2}}~.
\end{align}
In order to proceed further and to relate this velocity with that of escape velocity from the surface of rotating star, we need to impose the following boundary condition --- at large distance the escape velocity should tend to zero, i.e., as $r\rightarrow \infty$, $v^{2}\rightarrow 0$. Imposition of this boundary condition is not possible without an explicit expression for the metric functions. Thus we will now concentrate on the Hartle-Thorne metric, depicting the exterior vacuum geometry of a rotating star, in order to determine the escape velocity and hence obtaining the corresponding Buchdahl bound.  

In the absence of any exact exterior solution to an arbitrary rotating stellar object, we assume the exterior geometry to be described by the Hartle-Thorne metric, which is correct only upto quadratic order in the angular momentum and the corresponding spacetime metric reads \cite{Hartle:1968si}, 
\begin{align}\label{Hartle_Thorne}
ds^{2}&=-f(r)\Big[1+2\chi^{2}\left(j_{0}+j_{2}P_{2}\right)\Big]dt^{2}+\frac{1+\frac{2\chi^{2}\left(m_{0}+m_{2}P_{2}\right)}{r-2\mathcal{M}(r)}}{1-\frac{2\mathcal{M}(r)}{r}}dr^{2}
\nonumber
\\
&\qquad +r^{2}\Big[1+2\chi^{2}\left(v_{2}-j_{2}\right)P_{2}\Big]\Big[d\theta^{2}+\sin^{2}\theta\left(d\phi-\chi\Omega dt\right)^{2}\Big]~.
\end{align}
Here, we have kept terms upto second order in the dimensionless spin parameter $\chi\equiv (a/M)$, where $a$ is the rotation parameter of the compact object i.e., we have ignored terms $\mathcal{O}(\chi^{3})$, and the above metric has the following radial dependences \cite{Hartle:1968si}, 
\begin{align}
\mathcal{M}(r)&=M\,,\quad f(r)=1-\frac{2M}{r}\,,\qquad \Omega(r)=\frac{2M^{2}}{r^{3}}
\label{first_HT}
\\
m_{0}(r)&=M\left(\delta m-\frac{M^{3}}{r^{3}}\right)~,
\\
j_{0}(r)&=-\frac{M\delta m}{r-2M}+\frac{M^{4}}{r^{3}(r-2M)}~,
\label{j0}
\\
m_{2}(r)&=\frac{M^{3}}{r^{4}}(5M-r)(r-2M)
\nonumber
\\
&\qquad +\frac{\delta q}{2M^{2}r}\Big[M(M-r)(3r^{2}-2M^{2}-6Mr)+3r^{2}(r-2M)^{2}\tanh^{-1}\left(\frac{M}{r-M}\right) \Big]~,
\\
j_{2}(r)&=\frac{M^{3}(M+r)}{r^{4}}
\nonumber
\\
&+\frac{\delta q}{2M^{2}r(r-2M)}\Big[M(r-M)(3r^{2}-2M^{2}-6Mr)-3r^{2}(r-2M)^{2}\tanh^{-1}\left(\frac{M}{r-M}\right)\Big]~,
\label{j2}
\\
v_{2}(r)&=-\frac{M^{4}}{r^{4}}+\frac{\delta q}{M}\Big[3(r-M)\tanh^{-1}\left(\frac{M}{r-M}\right)-\frac{M(M^{2}+3r^{2}-6Mr)}{r(r-2M)}\Big]~,
\label{Last_HT}
\end{align}
where, $P_{2}=(3\cos^{2}\theta-1)/2$ is a Legendre polynomial, $M$ is the monopole term among the mass moments, $\delta m$ is the dimensionless correction to the mass moment, i.e., $\delta m\equiv(\delta M/M)$ and $\delta q$ is the dimensionless correction to the quadrupole moment, defined by $\delta q=(\delta Q/M^{3})$, both appearing at $\mathcal{O}(\chi^{2})$. Thus on a $\theta=\textrm{constant}=\theta_{0}$ surface, by comparison with \ref{gen_metric}, we obtain the following expressions for the corresponding metric elements, 
\begin{align}
e^{2\nu}-\omega^{2}e^{2\psi}&=f(r)\Big[1+2\chi^{2}\left(j_{0}+j_{2}P_{2}\right)\Big]~,
\\
e^{2\mu_{2}}&=\left[1+\frac{2\chi^{2}\left(m_{0}+m_{2}P_{2}\right)}{r-2\mathcal{M}(r)}\right]\left(1-\frac{2\mathcal{M}(r)}{r}\right)^{-1}~,
\\
\omega e^{2\psi}&=\chi \Omega \Big[1+2\chi^{2}\left(v_{2}-j_{2}\right)P_{2}\Big]r^{2}\sin^{2}\theta~,
\\
e^{2\psi}&=\Big[1+2\chi^{2}\left(v_{2}-j_{2}\right)P_{2}\Big]r^{2}\sin^{2}\theta~.
\end{align}
Thus it follows that, $\omega=\chi \Omega$, $e^{2\nu}=f(r)\left[1+2\chi^{2}\left(j_{0}+j_{2}P_{2}\right)\right]+\Omega^{2}\chi^{2}r^{2}\sin^{2}\theta$, and $P_{2}=(3\cos^{2}\theta_{0}-1)/2$. These results will be used in the subsequent sections for two specific values of $\theta_{0}$, namely $\theta_{0}=0$ and $\theta_{0}=(\pi/2)$, corresponding to the axial and the equatorial planes, respectively.

\subsection{Equatorial Buchdahl bound of a slowly rotating star}

\begin{figure}
\centering
\begin{subfigure}[t]{0.45\textwidth}
\centering
\includegraphics[width=\textwidth]{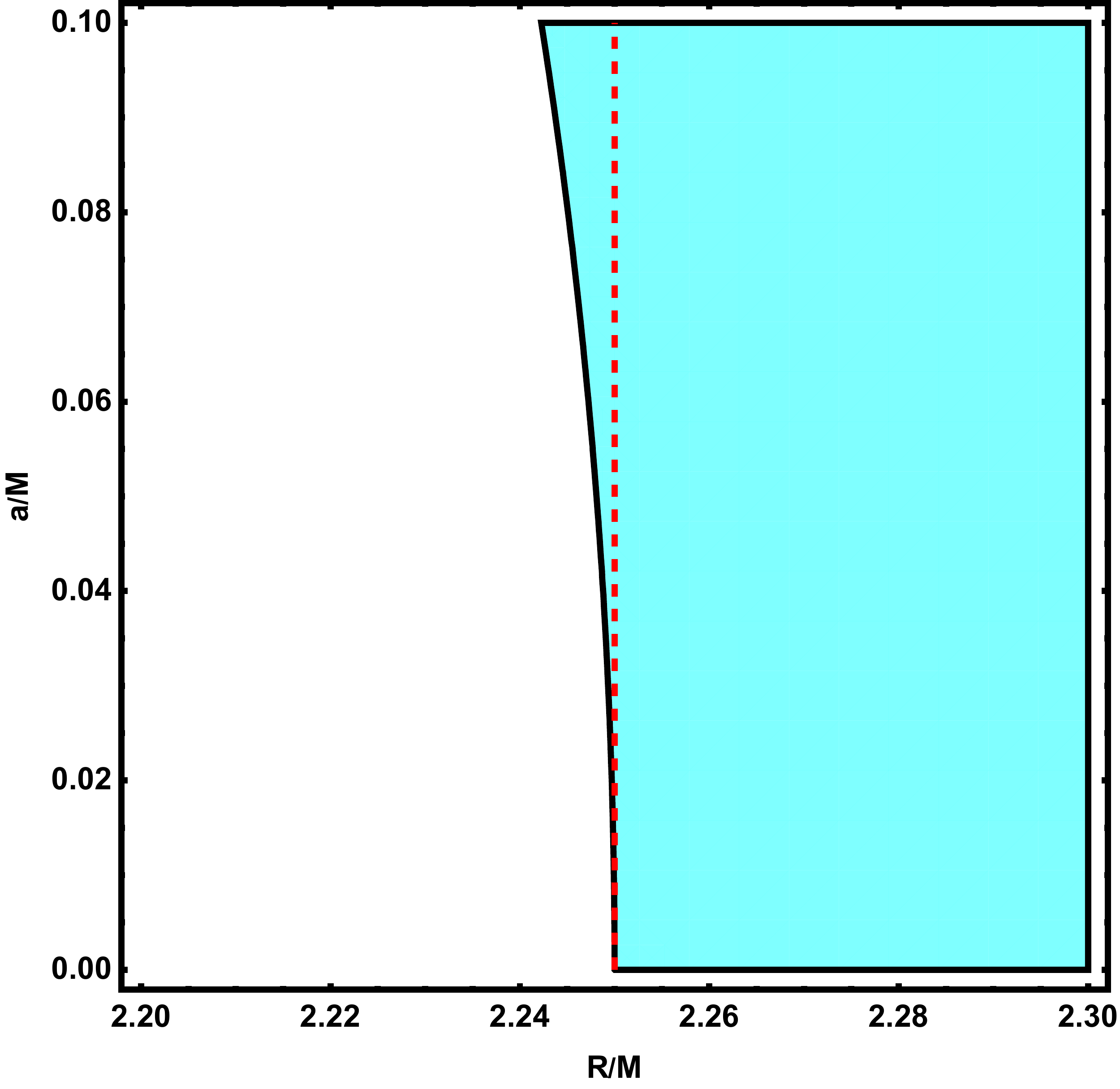}
\caption{The variation of the Buchdahl bound with the dimensionless rotation parameter $(a/M)$ has been presented, for the following choices for the dimensionless quantities: $\delta m=0.1=\delta q$.}
\end{subfigure}\hspace{1em} 
\begin{subfigure}[t]{0.45\textwidth}
\centering
\includegraphics[width=\textwidth]{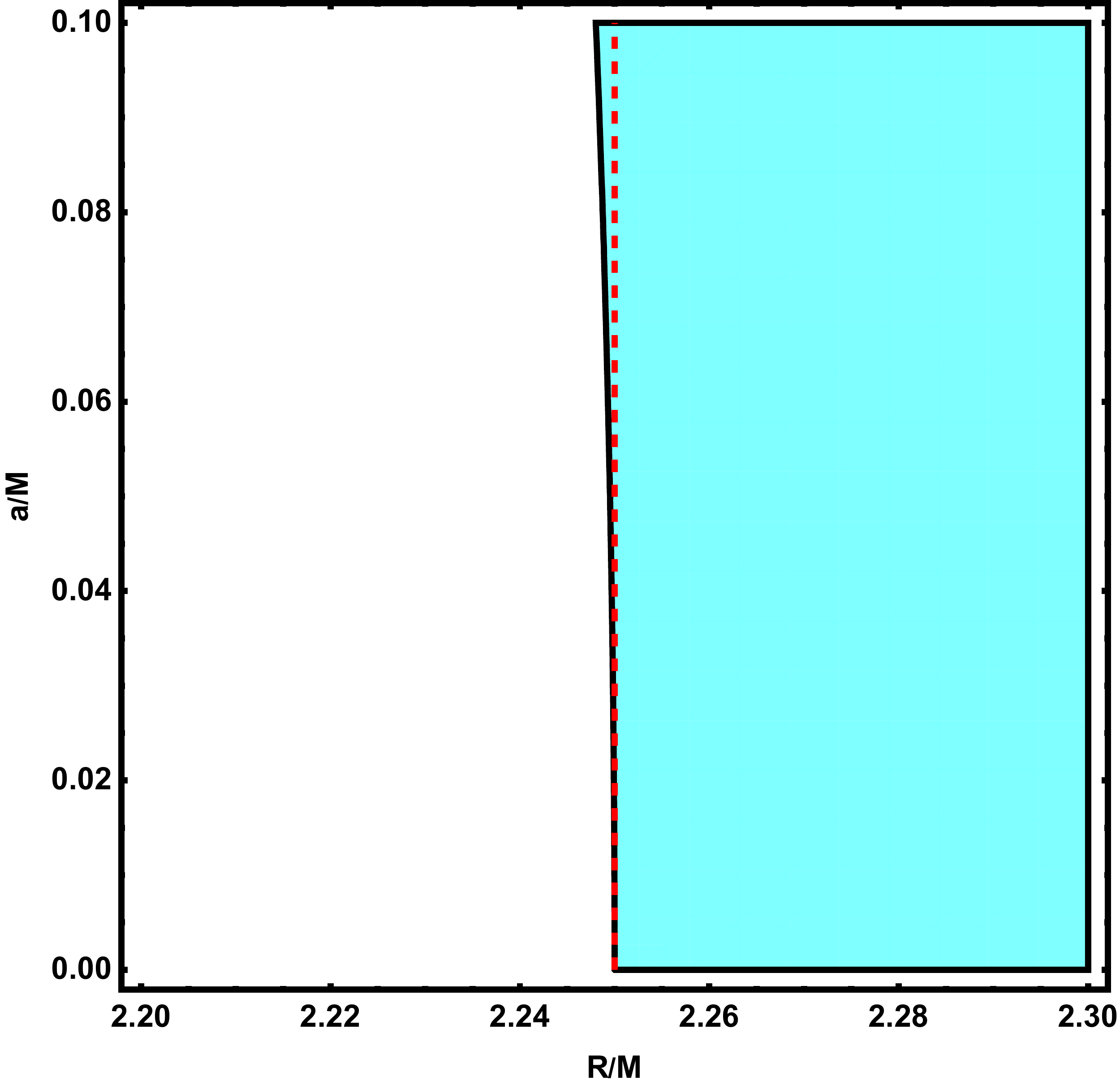}
\caption{The Buchdahl bound has been plotted against the dimensionless rotation parameter $(a/M)$, with the following choices for the dimensionless quantities: $\delta m=0.4=\delta q$.}
\end{subfigure}
\begin{subfigure}[t]{0.45\textwidth}
\centering
\includegraphics[width=\textwidth]{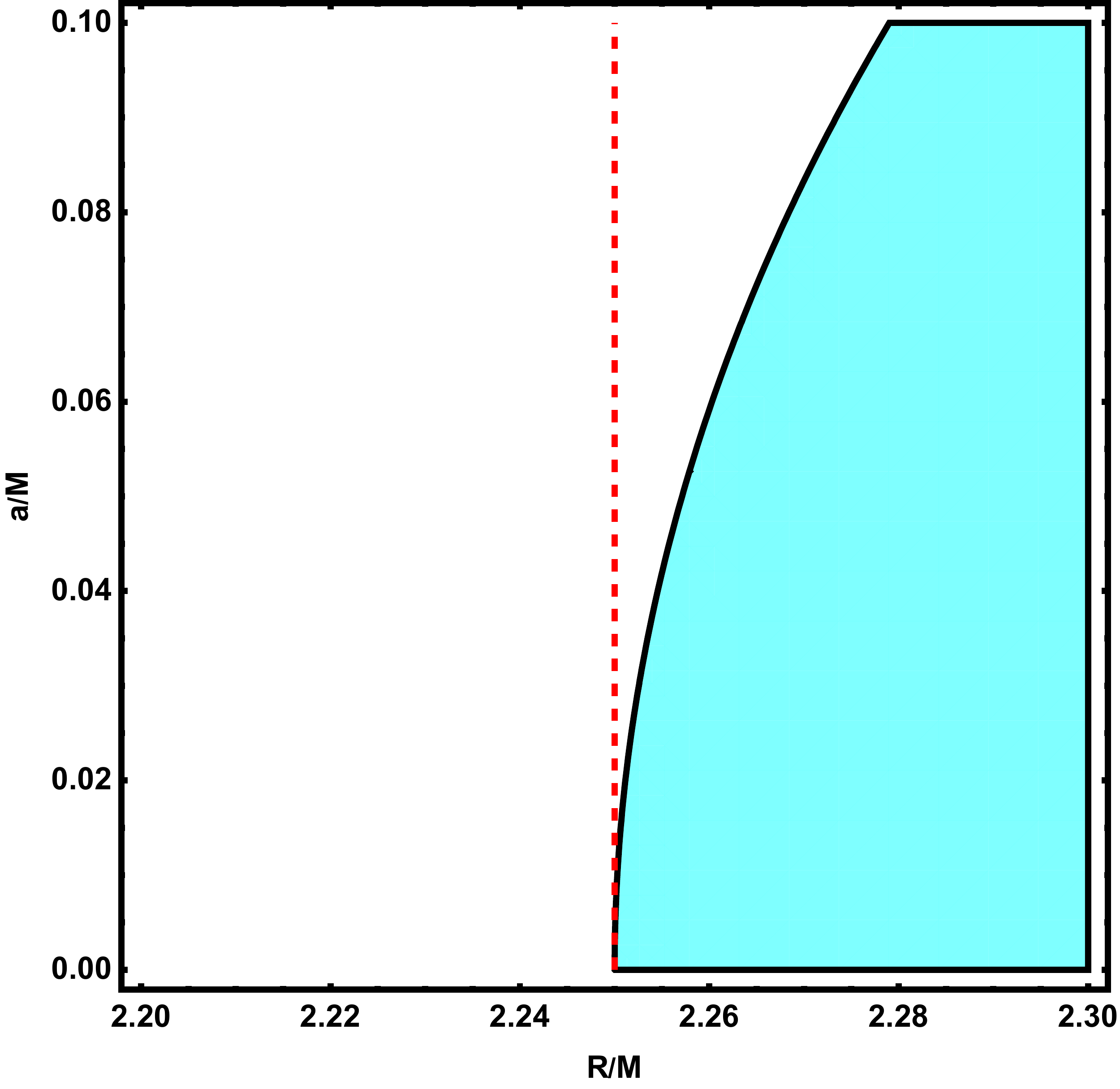}
\caption{The variation of the Buchdahl bound has been depicted against the dimensionless rotation parameter $(a/M)$, for the following choices: $\delta m=2=\delta q$.}
\label{axial}
\end{subfigure}
\caption{The region satisfying the inequality in \ref{inequality_equatorial}, associated with the Buchdahl bound on the equatorial plane, has been presented in the $\{(R/M),(a/M)\}$ plane, for different choices of the dimensionless parameters $\delta m=(\delta M/M)$ and $\delta q=(\delta Q/M^{3})$. The red dashed curve depicts the Buchdahl bound $(R/M)=(9/4)$ for static and spherically symmetric star. See text for more discussion.}
\label{equatorial}
\end{figure}

In this section, we will consider the Buchdahl bound on the equatorial plane of a slowly rotating star, whose exterior is described by the Hartle-Thorne metric, given by \ref{Hartle_Thorne}. On the equatorial plane, $P_{2}=-(1/2)$, and hence the non-zero metric elements associated with the exterior geometry reads, 
\begin{align}
e^{4\nu}&=\left\{f(r)\left[1+2\chi^{2}\left(j_{0}-\frac{1}{2}j_{2}\right)\right]+\Omega^{2}\chi^{2}r^{2}\right\}^{2}
\\
e^{2\nu}-\omega^{2}e^{2\psi}&=f(r)\Big[1+2\chi^{2}\left(j_{0}-\frac{1}{2}j_{2}\right)\Big]~;
\qquad 
\omega=\chi\Omega~,
\end{align}
where, the radial dependence of all the quantities appearing in the above equations have been presented in \ref{first_HT} to \ref{Last_HT}. Using these metric elements, which are functions of the radial coordinate alone, the expression for velocity of a particle on the equatorial plane, presented in \ref{escape_rotate}, takes the form,
\begin{align}
v^{2}=1-\frac{\left\{f(r)\left[1+2\chi^{2}\left(j_{0}-\frac{1}{2}j_{2}\right)\right]+\Omega^{2}\chi^{2}r^{2}\right\}^{2}}{f(r)\Big[1+2\chi^{2}\left(j_{0}-\frac{1}{2}j_{2}\right)\Big]\left(E-\chi \Omega L \right)^{2}}~.
\end{align}
If we now impose the condition that as $r\rightarrow \infty$, $v^{2}\rightarrow 0$, then we must have, $E=1$, necessity for an escape velocity. Also if we consider radial infall, then the angular momentum $L$ vanishes. Further expanding out all the terms in the above expression to $\mathcal{O}(\chi^{2})$, this leads to the following expression for the escape velocity on the equatorial plane from the surface of a rotating star with equatorial radius $R$ as,
\begin{align}
v_{\rm escape}^{2}=1-\left(1-\frac{2M}{R}\right)\Bigg[1+2\chi^{2}\left\{j_{0}(R)-\frac{1}{2}j_{2}(R)+\frac{R^{2}\Omega^{2}(R)}{\left(1-\frac{2M}{R}\right)}\right\}\Bigg]~,
\end{align}
where, $j_{0}$, $j_{2}$ and $\Omega$ are given by, \ref{first_HT}, \ref{j0} and \ref{j2}, respectively. If this radius $R$ has to represent the Buchdahl radius, our previous analysis suggests that the escape velocity should attain its limiting value $(8/9)$. Thus, in order to find out the Buchdahl limit on the equatorial plane, we impose the condition that $v_{\rm escape}^{2}\leq (8/9)$, and then we obtain the following inequality, 
\begin{align}\label{inequality_equatorial}
9\left(1-\frac{2M}{R}\right)\Bigg[1+2\chi^{2}\left\{j_{0}(R)-\frac{1}{2}j_{2}(R)+\frac{R^{2}\Omega^{2}(R)}{\left(1-\frac{2M}{R}\right)}\right\}\Bigg]\geq 1~.
\end{align}
In general, the above inequality is complicated in its dependence on the radial coordinate $R$, and hence we depict the same by plotting the region in the $\{(R/M),(a/M)\}$ plane, which satisfies the above inequality. This has been presented in \ref{equatorial}. As evident, for $a=0$, i.e., for zero rotation, the above inequality reduces to $(M/R)\leq (4/9)$, the expression for spherical stars, as expected. More importantly, the correction to the Buchdahl bound appears at the second order in the rotation parameter, implying for slowly rotating compact objects the modifications to the compactness ratio of (4/9) is small. Intriguingly, on the equatorial plane the radius of the star corresponding to the limiting escape velocity is smaller than the non-rotating counterpart, for smaller choices of the dimensionless mass and quadrupole moment deformation parameters $\delta m$ and $\delta q$, as evident by comparing the shaded region with the dashed line in the first two plots in \ref{equatorial}. However, for a larger dimensionless deformation parameters, as seen in the last plot of \ref{equatorial}, the limiting radius becomes larger. Thus rotation can make a star more compact on the equatorial plane for small mass and quadrupolar deformation, while for large deformation, the star becomes less compact on the equatorial plane. Note that the choice for these dimensionless parameters, namely $\delta m$ and $\delta q$ arise from the internal structure of the slowly rotating star and depends on the equation of state of the matter inside. For different equations of state and varied assumptions about the nature of the fluid, these dimensionless parameters can take values ranging from $\sim \mathcal{O}(0.3)$ \cite{Berti:2004ny} to $\sim \mathcal{O}(4)$ \cite{2021EPJC...81..698P}. We have chosen three sample values within this range to present the overall behaviour of the Buchdahl bound with the dimensionless rotation parameter $(a/M)$. In what follows we will adopt the same strategy for determining the variation of the Buchdahl bound with the dimensionless rotation parameter $(a/M)$, as well as the dimensionless mass and quadrupolar deformation along the axis of the slowly rotating object. 

\subsection{Axial Buchdahl bound on slowly rotating star}

\begin{figure}
\centering
\begin{subfigure}[t]{0.45\textwidth}
\centering
\includegraphics[width=\textwidth]{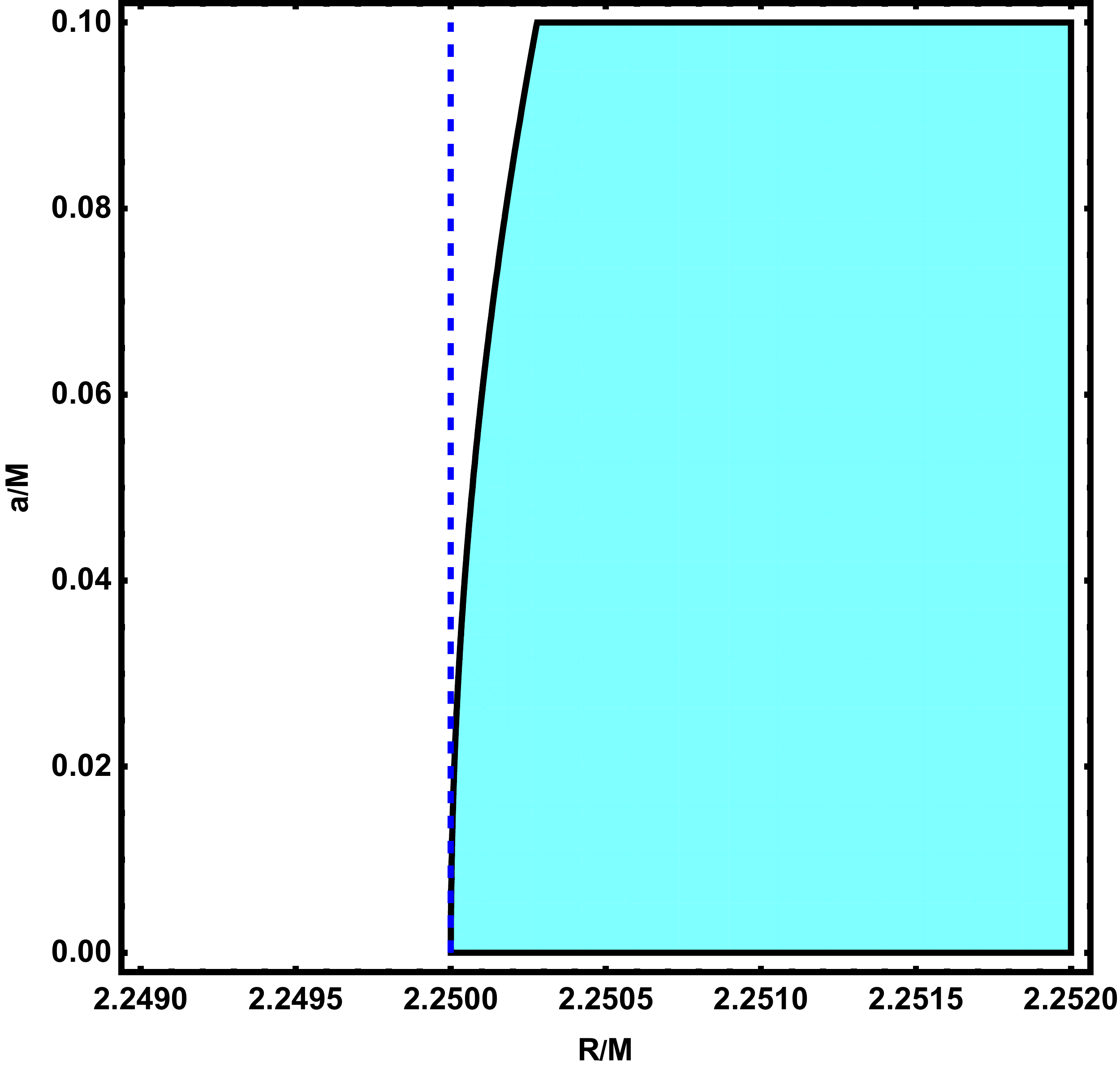}
\caption{The variation of axial Buchdahl bound with the dimensionless rotation parameter $(a/M)$ has been presented for the following choices of the dimensionless parameters: $\delta m=0.1=\delta q$.}
\end{subfigure}\hspace{1em} 
\begin{subfigure}[t]{0.45\textwidth}
\centering
\includegraphics[width=\textwidth]{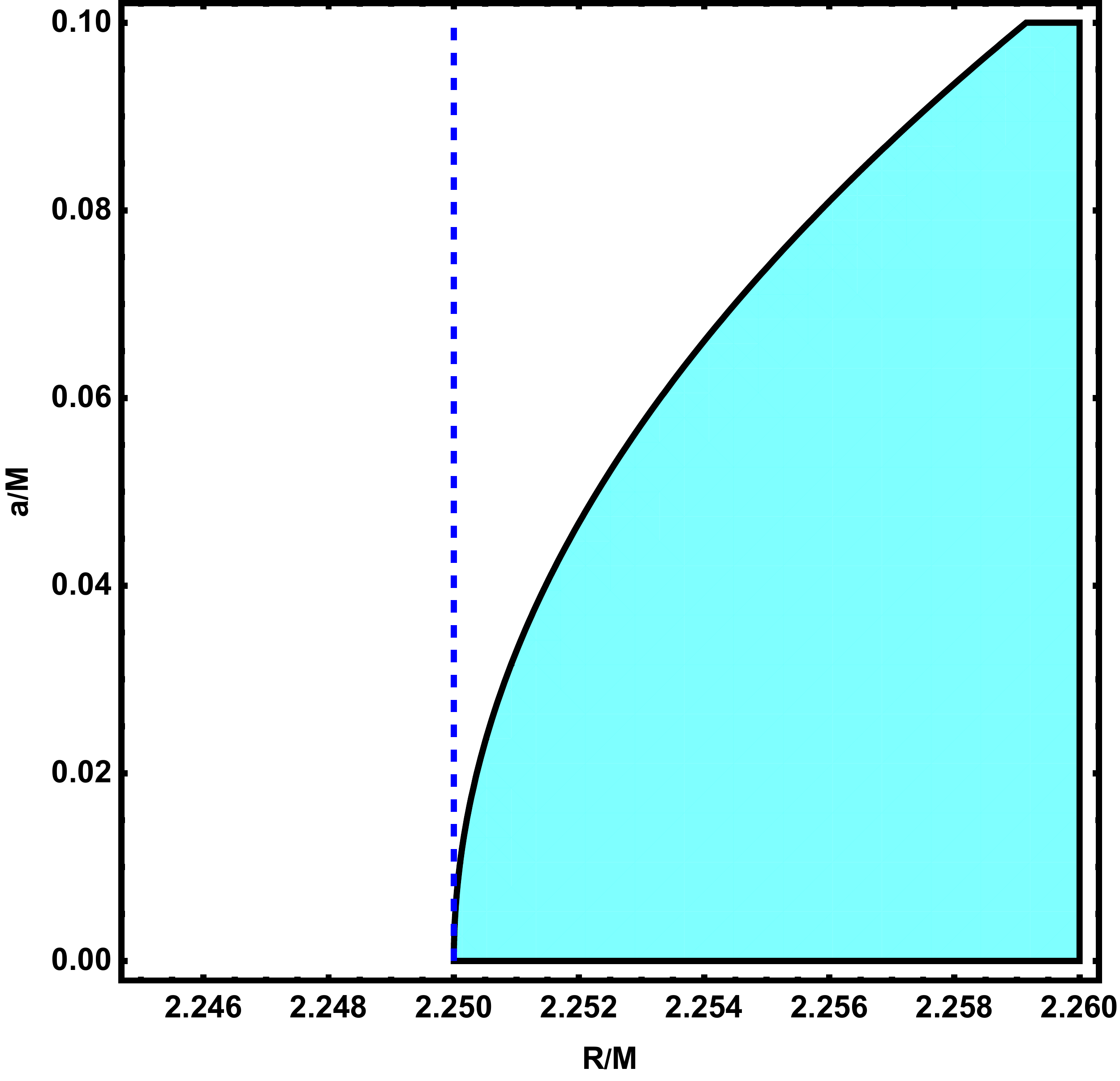}
\caption{The axial Buchdahl bound has been plotted against $(a/M)$, the dimensionless rotation parameter, for the following choices of the dimensionless deformation parameters $\delta m=0.4=\delta q$.}
\end{subfigure}
\begin{subfigure}[t]{0.45\textwidth}
\centering
\includegraphics[width=\textwidth]{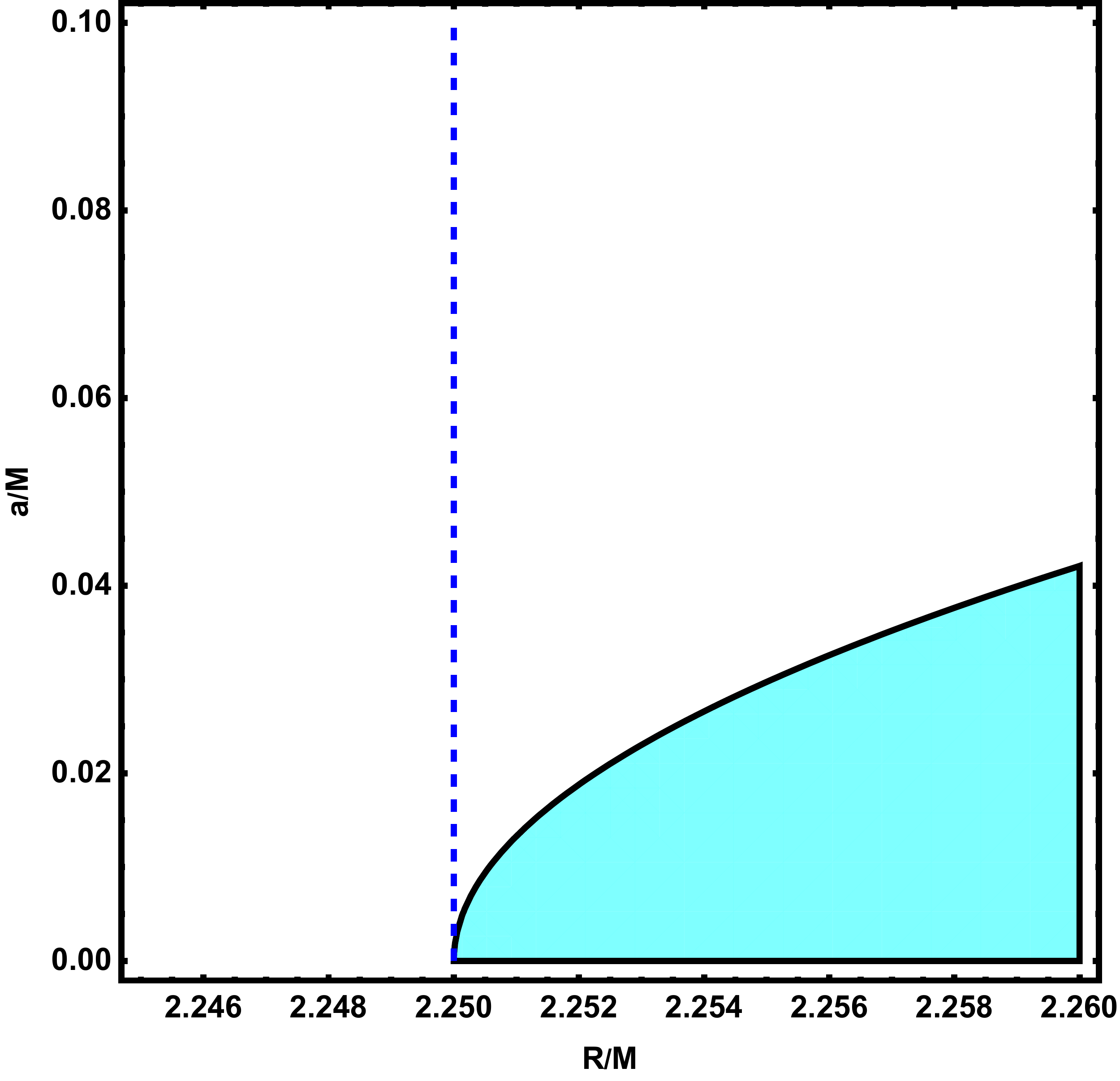}
\caption{The variation of the Buchdahl bound with the  dimensionless rotation parameter $(a/M)$ has been presented for $\delta m=2=\delta q$.}
\end{subfigure}
\caption{The region satisfying the inequality in \ref{axial_inequality}, associated with the Buchdahl bound along the axis, has been presented in the $\{(R/M),(a/M)\}$ plane. The blue dashed curve depicts the Buchdahl bound $(R/M)=(9/4)$ for static and spherically symmetric star. As evident, the radius at which the escape velocity becomes (8/9) on the axis, gets larger compared to the Buchdahl bound of static and spherically symmetric star, as the rotation parameter becomes larger. This enhancement of the Buchdahl radius becomes more prominent as the dimenionless deformation parameters become larger.}
\label{axial}
\end{figure}

In this section we discuss the radial motion of a particle, barely reaching infinity, escaping along the axis of rotation of the rotating star. This motion corresponds to $\theta=0$, along with the conserved energy and angular momentum being, $E=1$ and $L=0$, respectively, characteristics of the radial motion. Thus the relevant combinations of the metric elements, appropriate for the problem at hand, along the axis, become, 
\begin{align}
e^{4\nu}&=\left\{f(r)\left[1+2\chi^{2}\left(j_{0}+j_{2}\right)\right]+\Omega^{2}\chi^{2}r^{2}\right\}^{2}
\\
e^{2\nu}-\omega^{2}e^{2\psi}&=f(r)\Big[1+2\chi^{2}\left(j_{0}+j_{2}\right)\Big]~;
\qquad 
\omega=\chi\Omega~.
\end{align}
The radial dependence of all the quantities appearing in the above equations, namely $\Omega$, $j_{0}$ and $j_{2}$ have been presented in \ref{first_HT}, \ref{j0} and \ref{j2}, respectively. Using these radial functions, the escape velocity from the axis of a rotating Buchdahl star  becomes,
\begin{align}
v^{2}_{\rm escape}=1-\frac{\left\{f(R)\left[1+2\chi^{2}\left\{j_{0}(R)+j_{2}(R)\right\}\right]+\Omega^{2}(R)\chi^{2}R^{2}\right\}^{2}}{f(R)\Big[1+2\chi^{2}\left\{j_{0}(R)+j_{2}(R)\right\}\Big]}~,
\end{align}
where, $R$ is the radius of the rotating star at the axis of rotation. Expanding the above expression for the escape velocity in the dimensionless spin parameter and keeping terms $\mathcal{O}(\chi^{2})$, we obtain, 
\begin{align}
v^{2}_{\rm escape}=1-\left(1-\frac{2M}{R}\right)\Bigg[1+2\chi^{2}\left\{j_{0}(R)+j_{2}(R)\right\} \Bigg]~.
\end{align}
In order to find out the corresponding axial Buchdahl bound, we impose the condition, $v_{\rm escape}^{2}\leq (8/9)$, then we arrive at the following inequality, 
\begin{align}\label{axial_inequality}
9\left(1-\frac{2M}{R}\right)\Bigg[1+2\chi^{2}\left\{j_{0}(R)+j_{2}(R)\right\} \Bigg]\geq 1~.
\end{align}
As in the case of equatorial plane, here also it is not possible to get a closed form expression for the Buchdahl bound from the above inequality. This is why we have depicted the region in the $\{(R/M),(a/M)\}$ plane, which satisfies the above inequality, in \ref{axial}. As expected, for zero rotation, the Buchdahl bound reduces to $(M/R)\leq (4/9)$. However, for non-zero values of the rotation parameter, the limiting radius along the axis of the Buchdahl star increases and hence the compactness ratio decreases. It is worthwhile to mention that, unlike the Buchdahl radius on the equatorial plane, which first decreases and then increases with the increase of the dimensionless deformation parameters $\delta m=(\delta M/M)$ and $\delta q=(\delta Q/M^{3})$, on the axis the Buchdahl radius always increases. Also the departure of the radius of the rotating Buchdahl star from the spherically symmetric case is much smaller along the axis, in comparison to the equatorial limiting radius. We would like to emphasize that the above result is correct upto $\mathcal{O}(\chi^{2})$. Nonetheless, the above analysis provides the first ever bound on the compactness ratio of a slowly rotating Buchdahl star, using universal properties associated with the escape velocity.  

\section{Universality of gravitational field energy for Buchdahl sphere}\label{univ_grav_energy_buchdahl}

In addition to the universality of the escape velocity and the gravitational potential, as we will demonstrate, there is a universality of the gravitational field energy as well, again associated with the Buchdahl sphere. For this purpose, as in the earlier scenarios involving escape velocity, we consider first a spherically symmetric star of radius $r_{0}$ in four dimensional general relativity, whose gravitational field energy is given by \cite{Chakraborty:2015kva},
\begin{align}\label{grav_mass}
E_{\rm grav}=r_{0}\left(1-\sqrt{1-\frac{2M}{r_{0}}}\right)-M~,
\end{align}
which effectively comes from the Brown-York prescription. In essence, we subtract the energy content arising out of the mass distribution from the total energy of the system, obtained by the Brown-York prescription, since this subtracted amount must arise from the energy in the gravitational field itself and hence is appropriate to refer to as the gravitational field energy. If the star is now considered to be a Buchdahl sphere, with its radius $r_{0}$ coinciding with the radius $r_{\rm B}$ of the Buchdahl sphere, such that, $(2M/r_{\rm B})=(8/9)$, the gravitational field energy presented in \ref{grav_mass} becomes, 
\begin{align}
E_{\rm grav}^{\rm Buchdahl}=\frac{2}{9}r_{\rm B}~.
\end{align}
On the other hand, the non-gravitational energy associated with the spherical star in four dimensional general relativity is simply given by, $E_{\rm non-grav}=M$. The non-gravitational energy is precisely the one arising out of the matter distribution and hence in the present situation this is simply given by the mass of the star, which for the Buchdahl sphere reduces to, $E_{\rm non-grav}^{\rm Buchdahl}=(4/9)r_{\rm B}$. Therefore, it follows that the ratio of the gravitational and non-gravitational energy associated with the Buchdahl sphere becomes, 
\begin{align}
\frac{E_{\rm grav}^{\rm Buchdahl}}{E_{\rm non-grav}^{\rm Buchdahl}}=\frac{1}{2}~.
\end{align}
To see that whether this ratio is indeed universal, we now consider the case of a charged shell with radius $r_{0}$, again in the context of four dimensional general relativity. For which, the gravitational field energy becomes \cite{Chakraborty:2015kva}, 
\begin{align}\label{grav_en_charge}
E_{\rm grav}(r_{0})=r_{0}\left[1-\sqrt{1-\frac{2M}{r_{0}}+\frac{Q^{2}}{r_{0}^{2}}}\right]-\frac{r_{0}}{2}\left(\frac{2M}{r_{0}}-\frac{Q^{2}}{r_{0}^{2}}\right)~.
\end{align}
Here, the first term corresponds to the Brown-York energy within a sphere of radius $r_{0}$, larger than the radius of the spherical shell and the second term is the non-gravitational energy, due to the mass and the electric field energy within this radius $r_{0}$. Note that for $Q=0$, the above expression for gravitational field energy of a charged shell reduces to the gravitational field energy for a massive shell, as depicted in \ref{grav_mass}. On the charged Buchdahl shell, $r_{0}=r_{\rm B}$, where the radius $r_{\rm B}$ satisfies the following equation,
\begin{align}
\frac{2M}{r_{\rm B}}-\frac{Q^{2}}{r_{\rm B}^{2}}=\frac{8}{9}~,
\end{align}
and hence the gravitational field energy for the charged Buchdahl shell from \ref{grav_en_charge} becomes,
\begin{align}
E_{\rm grav}^{\rm Buchdahl}=r_{\rm B}\left[1-\sqrt{1-\frac{8}{9}}\right]-\frac{4r_{\rm B}}{9}=\frac{2}{9}r_{\rm B}~.
\end{align}
Further, for the spherically symmetric charged shell with radius $r_{0}$, the non-gravitational energy, arising from the mass and the electric field energy, takes the following form \cite{Chakraborty:2015kva},
\begin{align}
E_{\rm non-grav}(r_{0})=\frac{r_{0}}{2}\left(\frac{2M}{r_{0}}-\frac{Q^{2}}{r_{0}^{2}}\right)~,
\end{align}
therefore, it follows that on the Buchdahl sphere the non-gravitational energy becomes,
\begin{align}
E_{\rm non-grav}^{\rm Buchdahl}=\frac{4r_{\rm B}}{9}~.
\end{align}
Therefore, the ratio of gravitational and non-gravitational field energy on the charged Buchdahl shell in four dimensional general relativity becomes, 
\begin{align}
\frac{E_{\rm grav}^{\rm Buchdahl}}{E_{\rm non-grav}^{\rm Buchdahl}}=\frac{1}{2}~. 
\end{align}
This shows that irrespective of the presence of electrically charged or neutral matter content of spherically symmetric Buchdahl sphere, the ratio of the gravitational and the non-gravitational energy is universal for four dimensional general relativity and has a value equal to $(1/2)$. 

Let us now explore if this universality is respected for generic pure Lovelock theories as well, akin to the case of escape velocity from the Buchdahl sphere, whose universality transcends general relativity and holds true in pure Lovelock theories as well. The gravitational field energy for a sphere with radius $r_{0}$ in pure Lovelock theory can be computed following the Brown-York prescription, which yields \cite{Chakraborty:2015kva},
\begin{align}\label{grav_mass_love}
E_{\rm grav}(r_{0})=r_{0}^{(d-2N-1)/N}\left(1-\sqrt{1-\frac{2M^{1/N}}{r_{0}^{(d-2N-1)/N}}}\right)-M^{1/N}~,
\end{align}
where the first term corresponds to the Brown-York energy and the second one is the contribution from the mass. On the Buchdahl sphere with radius $r_{\rm B}$, we have the following relation between $r_{\rm B}$ and the mass of the stellar configuration: $r_{\rm B}^{(d-2N-1)/N}=\{(d-1)^{2}/2N(d-N-1)\}M^{1/N}$, such that, the gravitational field energy of the Buchdahl sphere from \ref{grav_mass_love} becomes,
\begin{align}
E^{\rm buchdahl}_{\rm grav}&=r_{\rm B}^{(d-2N-1)/N}\left[\left(1-\sqrt{1-\frac{4N(d-N-1)}{(d-1)^{2}}}\right)-\frac{2N(d-N-1)}{(d-1)^{2}}\right]
\nonumber
\\
&=2\alpha r_{\rm B}^{(d-2N-1)/N}\left[\frac{\left(1-\sqrt{1-4\alpha}\right)}{2\alpha}-1\right]~;\qquad \alpha=\frac{N(d-N-1)}{(d-1)^{2}}~.
\end{align}
While, the non-gravitational energy can be determined by using the fact that, on the black hole horizon both the gravitational field energy and the non-gravitational energy takes equal values. Such that, $E_{\rm grav}^{\rm hor}=M^{1/N}=E_{\rm non-grav}$ on the horizon $r_{\rm h}$, satisfying $r_{\rm h}^{(d-2N-1)/N}=2M^{1/N}$. Thus, with the expression for $M^{1/N}$ in terms of the Buchdahl radius of a spherically symmetric star in pure Lovelock gravity we obtain,
\begin{align}
\frac{E_{\rm grav}^{\rm buchdahl}}{E_{\rm non-grav}^{\rm buchdahl}}=\frac{\left(1-\sqrt{1-4\alpha}\right)}{2\alpha}-1~.
\end{align}
For, four dimensional general relativity, we have, $d=4$ and $N=1$, yielding $\alpha=(2/9)$ and then the above ratio reduces to $(1/2)$, as one can immediately verify. Moreover, even for the charged Buchdahl sphere in pure Lovelock gravity, one can also verify that the above ratio is indeed universal. Thus the ratio of gravitational and non-gravitational energy is universal for Buchdahl sphere irrespective of whether the sphere involves electric charge or not and it holds not only in general relativity but also for pure Lovelock theories of gravity. This further corroborates the fact that the results holding for general relativity also generalizes in the context of pure Lovelock gravity. Moreover, for a given $N$, if we choose $d=3N+1$, then also, we obtain, $\alpha=(2/9)$ and hence we obtain the maximal value of $\beta=(1/2)$. This again illustrates the fact that the pure Lovelock gravity of order $N$ in $d=3N+1$ dimensions has properties identical to that of four dimensional general relativity. 

\section{Universality of the photon sphere}

In this final section, we will demonstrate another universal property of the Buchdahl sphere, namely the ratio of the radius of the Buchdahl sphere to that of the photon sphere in the spherically symmetric context. As a first step towards exploring this universality consider the radius of the photon sphere in static and spherically symmetric geometry of the Schwarzschild spacetime, which becomes, $r_{\rm ph}=3M$, such that the ratio of the photon sphere and the Buchdahl sphere becomes, 
\begin{align}
\frac{r_{\rm ph}}{r_{\rm buchdahl}}=\frac{3M}{(9M/4)}=\frac{4}{3}~.
\end{align}
For the case of Reissner-Nordstr\"{o}m spacetime, the photon sphere needs to be obtained by solving the algebraic equation $rf'=2f$, where we have, $f(r)=1-(2M/r)+(Q^{2}/r^{2})$. Expanding out the above algebraic equation, it follows that the photon sphere satisfies the equation, 
\begin{align}
r\left(\frac{2M}{r^{2}}-\frac{2Q^{2}}{r^{3}}\right)=2\left(1-\frac{2M}{r}+\frac{Q^{2}}{r^{2}}\right)~,
\end{align}
which is equivalent to, $r^{2}-3Mr+2Q^{2}=0$, with the following solution,
\begin{align}
r_{\rm ph}=\frac{3M+\sqrt{9M^{2}-8Q^{2}}}{2}=\frac{3M}{2}\left[1+\sqrt{1-\frac{8}{9}\frac{Q^{2}}{M^{2}}}\right]~.
\end{align}
Note that, in general the above algebraic equation will have two independent solutions, however, the solution within the black hole horizon is not being considered here. For the case of a charged shell, whose outside geometry is Reissner-Nordstr\"{o}m, the buchdahl limit is given by \cite{Giuliani:2007zza,Boehmer:2007gq},
\begin{align}
r_{\rm buchdahl}=\frac{9M}{8}\left[1+\sqrt{1-\frac{8}{9}\frac{Q^{2}}{M^{2}}}\right]~.
\end{align}
Therefore, the ratio of the radius of photon sphere and Buchdahl sphere involving the Maxwell charge is given by, 
\begin{align}
\frac{r_{\rm ph}}{r_{\rm buchdahl}}=\frac{4}{3}
\end{align}
Thus we observe that for four dimensional general relativity, the ratio of the radius of the photon sphere and the radius of the Buchdahl sphere is universal, i.e., irrespective of whether the star carries Maxwell charge or, not. Unfortunately, unlike the previous cases, the above universality cannot be generalized for pure Lovelock theories of gravity. This is because, the equation for the photon sphere in $N$th order pure Lovelock theories of gravity is a higher order algebraic equation, depending on the dimensions of the spacetime and the Lovelock order $N$, which is difficult to solve for. Thus the ratio of the photon sphere and the Buchdahl sphere cannot be arrived at in a closed form. Thus this universal feature of the Buchdahl sphere holds in four dimensional general relativity alone. 

\section{Concluding Remarks}

The Buchdahl sphere --- a stellar configuration having its radius equal to the Buchdahl limit, the limiting configuration beyond which the star will collapse to a black hole --- has several universal characters, which we have explored in this work. In particular, we have shown that the following properties of a Buchdahl sphere --- (a) the escape velocity from the Buchdahl sphere, which in turn is related to the gravitational potential on the Buchdahl sphere, (b) the ratio of quasi-local gravitational and non-gravitational energies associated with the Buchdahl sphere and (c) the ratio of the Buchdahl limit with that of the photon sphere --- are universal in general relativity, i.e., are independent of the presence of electrically neutral or, charged matter within the star. Intriguingly most of these characteristics transcend general relativity and holds for pure Lovelock theories of gravity as well. Using the universality of the escape velocity, we have also derived the Buchdahl limit of a slowly rotating stellar configuration in general relativity using Hartle-Thorne metric --- even though no direct computation of the same exists using the interior of a rotating stellar configuration --- for the first time in the literature. Intriguingly, the limiting compactness of a slowly rotating stellar configuration is more along the equator, than along the axis.  All of these results boil down to the static and spherically symmetric counterpart in appropriate limits.  

These universal features of the Buchdahl sphere, along with its compactness, which is of the same order as that of a black hole, implies that the Buchdahl sphere is expected to mimic almost all or most of the features of a black hole spacetime. For example, in terms of the gravitational potential on the horizon (on the Buchdahl limit), we obtain, $\Phi=0.5$ ($\Phi=4/9)$ in general relativity, which is universal irrespective of the presence of charge and rotation, and arises from the universality of the escape velocity in the respective situations, with $v_{\rm escape}^{2}=1$ ($v_{\rm escape}^{2}=8/9)$. Moreover, the gravitational redshift at the surface of the Buchdahl sphere always has the value $3$, irrespective of charge or, rotation. Since the Buchdahl sphere is as compact as a black hole, but having a non-null boundary, it can lead to interesting physical scenarios. For example, if a Buchdahl sphere is perturbed, the boundary condition at the surface of the sphere will be different from that of the black hole. In addition, the separation between the photon sphere and the Buchdahl sphere is $\mathcal{O}(M)$, and hence the ringdown spectrum of the Buchdahl sphere, under perturbation, will inherit echoes, which can be probed using future gravitational wave detectors. In this sense, the Buchdahl sphere is an astrophysically promising and interesting object. 

It should also be noted that all of these universal features follow entirely from the exterior solution, without any reference to the interior matter distribution and geometry, except for some general requirements. For example, the stellar material must obey the energy conditions and the energy density and pressure distribution must be finite at all points within the star. These universal features also result into the Buchdahl sphere being accessible to astrophysical observations. For instance, if an object  is observed, with its gravitational redshift at the surface being larger than $3$, it would strongly indicate that it is more compact that a Buchdahl sphere, and without any exotic matter support, is possibly a black hole. Thus the surface red shift could serve as a good practical way of distinguishing the Buchdahl sphere from a black hole or, for that matter from exotic compact objects.

Another interesting aspect associated with the Buchdahl sphere is the result that the extremal limit for a Buchdahl sphere is actually over extremal, relative to a black hole spacetime. For instance, in the case of a charged Buchdahl sphere, the extremal limit corresponds to $(Q^2/M^2)=(9/8)>1$. Thus it is worthwhile to ask, whether it will be possible to form an extremal charged BH by starting from a near extremal Buchdahl sphere with $1<(Q^2/M^2)<(9/8)$ and then throwing in neutral matter. This could perhaps be the only way to form an extremal black hole, if at all. Further, it is also known that non-extremal black holes cannot be extremalized by adiabatic test particle accretion, following which we wish to study whether a similar situation holds for Buchdahl sphere as well in a future work. In this connection, it is also of interest to probe whether a Buchdahl sphere can be over-extremalized and whether the results similar to the weak cosmic censorship conjecture for black holes hold good for Buchdahl sphere as well. We leave these issues for the future. 

\section*{Acknowledgements}

Research of S.C. is funded by the INSPIRE Faculty fellowship from DST, Government of India (Reg. No. DST/INSPIRE/04/2018/000893) and by the Start-Up Research Grant from SERB, DST, Government of India (Reg. No. SRG/2020/000409). ND wishes to thankfully acknowledge support of the CAS President's International Fellowship Initiative Grant No. 2020VMA0014 for support. 
\appendix
\labelformat{section}{Appendix #1} 
\labelformat{subsection}{Appendix #1}
\bibliography{Universal_Buchdahl_References}

\bibliographystyle{./utphys1}
\end{document}